\documentclass{ws-procs975x65}

\begin{document}

\title{GENERIC GRAVITY TESTS WITH THE DOUBLE PULSAR}

\author{NORBERT WEX$^*$ and MICHAEL KRAMER}

\address{
Max-Planck-Institut f\"ur Radioastronomie,\\
Auf dem H\"ugel 69, D-53121 Bonn, Germany\\
$^*$E-mail: wex@mpifr-bonn.mpg.de}

\begin{abstract}
Presently the double pulsar allows for the measurement of six post-Keplerian
parameters. In addition, its double-line nature gives access to the projected
semi-major axes of both orbits. We use this wealth of information to pose some
very general restrictions on a wide class of conservative and
semi-conservative theories of gravity.
\end{abstract}


\bodymatter\bigskip

\noindent
Among the most well-known uses of pulsars has been their role in tests of
theories of gravity, in particular in the experimental verification of general
relativity (GR) \cite{tay94b}. In 2003 a binary system was discovered where
both members were identified as radio pulsars. \cite{bdp+03, lbk+04}. Both
pulsars, now known as PSR J0737--3039A (23\,ms) and PSR J0737--3039B (2.8\,s),
$A$ and $B$ hereafter, orbit each other in 2.45 hours in a slightly eccentric
orbit. As a result, the system is not only the first and, thus far, only
double neutron star system where both neutron stars are visible as active
radio pulsars, but it is also the most relativistic binary pulsar known to
date. By measuring the radial motion of the two pulsars, we obtain a precise
measurement of the mass ratio $m_A/m_B$ \cite{dam07}. In addition, a total of
six post-Keplerian (PK) parameters has been measured. Five arise from four
different relativistic effects visible in pulsar timing (Ref.~\citen{ksm+06}),
while a sixth one can be determined from the effects of geodetic precession on
the observed radio eclipses (Ref.~\citen{bkk+08}). The measurement of the mass
ratio and six PK parameters, including the relativistic spin precession of
$B$, makes the double pulsar (DP) the most constrained binary pulsar
known. This allows for much more general statements about alternative theories
of gravity than it has been possible before the discovery of this system.

\subsection*{Testing conservative theories of gravity}

Will (Ref.~\citen{wil93}) and Damour and Taylor (Ref.~\citen{dt92}) have
generalized the Lagrangian of the post-Newtonian orbital dynamics to the
strong-field regime, for fully conservative theories of gravity. Therein,
strong-field effects (SFEs) in the orbital motion of a binary system that
consists of two compact bodies are accounted for by three SFE parameters:
${\cal G}$ (effective gravitational constant), $\varepsilon$, and
$\xi$. Possible SFEs in the spin--orbit interaction are described by two
coupling functions $\Gamma_A^B$ and $\Gamma_B^A$ \cite{dt92}.

In GR ${\cal G} = G$, $\varepsilon = 3$, $\xi = 1$, and $\Gamma_A^B =
\Gamma_B^A = 2G$. In alternative theories of gravity these quantities are
functions of the parameters of the theory and of the structure of each
body. For neutron stars these parameters can deviate significantly from their
values in GR, even if their weak-field limit agrees with GR \cite{wil93,
  de96a}.

The fact that in any Lorentz-invariant theory of gravity the DP gives access
to the mass ratio through its `double-line' nature (Ref.~\citen{dam07}) and
the inclination of the orbit with respect to the line of sight via the shape
of the Shapiro delay (Refs.~\citen{dt92, wil93}), allows us to determine
theory independent values for the effective gravitating masses of the system:
${\cal G} m_A = 1.339 \pm 0.003 \, GM_\odot$ and ${\cal G} m_B = 1.250 \pm
0.002 \, GM_\odot$, where $M_\odot$ is the mass of the Sun \cite{kw09}.

Since the PK parameters of the Damour--Deruelle timing model can be expressed
as a function of the Keplerian parameters, the effective gravitating masses,
and the SFE parameters (Ref.~\citen{dt92}), we can use the measured PK
parameters to generically constrain the SFE parameters for the DP system:
\cite{kw09} \\[1mm]
\begin{tabular}{lclll}
  $\dot\omega$ & $\Rightarrow$ 
               & $2\varepsilon - \xi$ 
               & = \quad $4.998 \pm 0.008$ & [0.2\%] \\
  $\gamma$     & $\Rightarrow$ 
               & ${\cal G}_{B0}/{\cal G} + {\cal K}_A^B$ 
               & = \quad $1.005 \pm 0.010$ & [1.0\%] \\
  $r$          & $\Rightarrow$ 
               & $({\cal G}_{B0}/{\cal G})(\varepsilon_{B0} + 1)$ 
               & = \quad $4.04 \pm 0.22$   & [5.4\%] \\
  $\Omega_B$   & $\Rightarrow$ 
               & $\Gamma_B^A / {\cal G}$
               & = \quad $1.88 \pm 0.26$   & [13\%]
\end{tabular}
\\[1mm] 
which is in full agreement with GR. Index $B0$ indicates the SFE parameter
for the interaction between the compact body $B$ and a photon. The parameter
${\cal K}_A^B$ accounts for a possible change of the moment of inertia of
pulsar $A$ due to the changing distance to the compact companion $B$. This
parameter is zero in GR.

We would like to emphasize that currently the DP is the only system that
allows for the test of the ‘effacement’ property of a spinning body, i.e.\ the
fact that the spin-precession of a body does not depend on its internal 
structure (up to a certain post-Newtonian level) \cite{th85, dam87, kv08}.

\subsection*{Testing preferred-frame effects}

The generic tests of the previous section can be extended to semi-conservative
gravity theories, i.e.\ the DP can even be used to derive general restrictions
on theories in which gravity is not boost invariant. Such theories predict
that the Universe's global matter distribution selects a preferred rest frame
for local gravitational physics. A binary pulsar that moves with respect to
this preferred frame would show characteristic variations in the eccentricity
and orientation of its orbit, which can be used to test strong-field
preferred-frame effects (PFEs) \cite{de92a}.

The existence of a preferred frame for gravity would lead to a characteristic
signature in the timing data of the DP, as PFEs impose periodic changes on the
orbital parameters with periods of 21.30 and 10.65 years \cite{wk07}.
Therefore the DP has the potential to uniquely identify PFEs, if present at a
measurable level. Moreover, in the absence of PFEs the DP can be used to
determine limits for the strong-field PFE parameters ${\alpha}_1^\ast$ and
$\alpha_2^\ast$, corresponding to the 21.30 and 10.65 year period
respectively, for nearly any direction in the sky. Using the published timing
data (Ref.~\citen{ksm+06}), one finds for a preferred frame at rest with
respect to the cosmic microwave background:\footnote{Here we assume that the
  kinetic PFE parameters are equal to one, as in GR. Limits for the full PFE
  parameters, including the kinetic contributions, can be found in
  Ref.~\citen{wk07}.}  $-0.3 < \alpha_1^\ast < 0.2$, $-0.6 < \alpha_2^\ast <
0.3$.  These limits are clearly less stringent than present limits for
$\alpha_1^\ast$ from small-eccentricity binary pulsars \cite{de92a,
  wex00}. However, these results are based on timing observations which span
only a small fraction of the PFE periods given above, and therefore show very
strong correlations between the PFE amplitudes and other orbital
parameters. Simulations show that these correlations will reduce considerably
once the advance of periastron approaches $180^\circ$. Hence, an analysis
including all the new data is expected to provide limits that are better by
about three orders of magnitude.  Furthermore, we would like to emphasize that
preferred-frame tests with the DP are qualitatively different from those in
pulsar white-dwarf binaries, since it probes effects that are only present in
the interaction of two compact bodies \cite{wil93, de96b, wk07}.


\begin{thebibliography}{10}

\bibitem{tay94b}
J.~H.~Taylor, {\it Binary pulsars and relativistic gravity}, 
  in: Les Prix Nobel, Norstedts Tryckeri, Stockholm, 1994, pp.~80.

\bibitem{bdp+03}
M.~Burgay, N.~D'Amico, A.~Possenti, R.~N.~Manchester, A.~G.~Lyne,
  B.~C.~Joshi, M.~A.~McLaughlin, M.~Kramer, J.~M.~Sarkissian,
  F.~Camilo, V.~Kalogera, C.~Kim and D.~R.~Lorimer, 
  {\it Nature}~{\bf 426} 531--533~(2003).

\bibitem{lbk+04}
A.~G. Lyne, M.~Burgay, M.~Kramer, A.~Possenti, R.~N. Manchester, F.~Camilo,
  M.~A. McLaughlin, D.~R. Lorimer, N.~D'Amico, B.~C. Joshi, J.~Reynolds
  and P.~C.~C. Freire, {\it Science}~{\bf  303} 1153--1157~(2004).

\bibitem{dam07}
T.~Damour, {\it Binary Systems as Test-beds of Gravity Theories}, 
  in:  M.~Colpi {\it et al.} (Eds.), A Century from Einstein Relativity: 
  Probing Gravity Theories in Binary Systems, 
  Springer, 2007, (arXiv:0704.0749).

\bibitem{ksm+06}
M.~Kramer, I.~H.~Stairs, R.~N.~Manchester, M.~A.~McLaughlin, 
  A.~G.~Lyne, R.~D.~Ferdman, M.~Burgay, D.~R.~Lorimer, A.~Possenti,
  N.~D'Amico, J.~M.~Sarkissian, G.~B.~Hobbs, J.~E.~Reynolds
  and P.~C.~C.~Freire, F.~Camilo, {\it Science}~{\bf  314} 97--102~(2006).

\bibitem{bkk+08}
R.~P.~Breton, V.~M.~Kaspi, M.~Kramer, M.~A.~McLaughlin, M.~Lyutikov,
  S.~M.~Ransom, I.~H.~Stairs, R.~D.~Ferdman, F.~Camilo and A.~Possenti,
  {\it Science}~{\bf  321} 104--107~(2008).

\bibitem{wil93}
C.~M.~Will, {\it Theory and Experiment in Gravitational Physics}, 
  Cambridge University Press, Cambridge, 1993.

\bibitem{dt92}
T.~Damour and J.~H.~Taylor, {\it Phys. Rev. D}~{\bf  45} 1840--1868~(1992).

\bibitem{de96a}
T.~Damour and G.~Esposito-Farese, {\it Phys. Rev. D}~{\bf  54} 
  1474--1491~(1996).

\bibitem{kw09}
M.~Kramer and N.~Wex, {\it Class. Quantum Grav.}~{\bf  26} 073001~(2009).

\bibitem{dam87}
T.~Damour, {\it The problem of motion in Newtonian and Einsteinian gravity}, 
  in: S.~Hawking and W.~Israel (Eds.), 300 Years of Gravitation, 
  Cambridge University Press, Cambridge, 1987, pp.~128.

\bibitem{th85}
K.~S.~Thorne and J.~B.~Hartle, {\it Phys. Rev. D}~{\bf 31} 1815--1837~(1985).

\bibitem{kv08}
S.~Kopeikin and I.~Vlasov, {\it The effacing principle in the post-Newtonian 
  Celestial Mechanics}, in: H.~Kleinert, R.T.~Jantzen and R.~Ruffini (Eds.), 
  Proc.\ 11th Marcel Grossmann Meeting, part C, 
  World Scientific, 2008, pp.~2475.

\bibitem{de92a}
T.~Damour and G.~Esposito-Far\`ese, 
  {\it Phys. Rev. D}~{\bf  46} 4128--4132~(1992).

\bibitem{wk07}
N.~Wex and M.~Kramer, {\it MNRAS}~{\bf  380} 455--465~(2007).

\bibitem{wex00}
N.~Wex, {\it Small-eccentricity binary pulsars and relativistic gravity},
  in: M.~Kramer, N.~Wex, and R.~Wielebinski (Eds.), 
  Pulsar Astronomy -- 2000 and Beyond, IAU Colloquium 177,
  Astronomical Society of the Pacific, San Francisco, 2000, pp.~113.

\bibitem{de96b}
T.~Damour and G.~Esposito-Far{\`e}se, 
  {\it Phys. Rev. D}~{\bf  53}, 5541--5578~(1996).

\end{thebibliography}
\end{document}